\documentclass[prd,superscriptaddress,amsfonts,amssymb,amsmath,showpacs,twocolumn]{revtex4-2}
\usepackage{bm}
\usepackage{amsfonts}
\usepackage{latexsym}
\usepackage[latin1]{inputenc}
\usepackage{graphicx}
\usepackage{amsmath}
\usepackage{palatino}
\usepackage{mathpazo}
\usepackage{textcomp}
\usepackage{float}
\usepackage{booktabs}
\usepackage{dcolumn}
\usepackage{multirow}
\usepackage{ragged2e}
\usepackage{hyperref}
\hypersetup{colorlinks,citecolor=blue}
\usepackage{amsmath}
\usepackage{xcolor}
\usepackage{orcidlink}
\usepackage[caption=false]{subfig}
\usepackage{commath}
\def\jnl@style{\it}
\def\aaref@jnl#1{{\jnl@style#1}}

\def\aaref@jnl#1{{\jnl@style#1}}

\def\aj{\aaref@jnl{AJ}}                   % Astronomical Journal
\def\apj{\aaref@jnl{ApJ}}                 % Astrophysical Journal
\def\apjl{\aaref@jnl{ApJ}}                % Astrophysical Journal, Letters
\def\apjs{\aaref@jnl{ApJS}}               % Astrophysical Journal, Supplement
\def\apss{\aaref@jnl{Ap\&SS}}             % Astrophysics and Space Science
\def\aap{\aaref@jnl{A\&A}}                % Astronomy and Astrophysics
\def\aapr{\aaref@jnl{A\&A~Rev.}}          % Astronomy and Astrophysics Reviews
\def\aaps{\aaref@jnl{A\&AS}}              % Astronomy and Astrophysics, Supplement
\def\mnras{\aaref@jnl{Mon.~Not.~Roy.~Astron.~Soc.}}             % Monthly Notices of the RAS
\def\prd{\aaref@jnl{Phys.~Rev.~D}}        % Physical Review D
\def\prc{\aaref@jnl{Phys.~Rev.~C}}  % Physical Review C
\def\prl{\aaref@jnl{Phys.~Rev.~Lett.}}    % Physical Review Letters
\def\qjras{\aaref@jnl{QJRAS}}             % Quarterly Journal of the RAS
\def\skytel{\aaref@jnl{S\&T}}             % Sky and Telescope
\def\ssr{\aaref@jnl{Space~Sci.~Rev.}}     % Space Science Reviews
\def\zap{\aaref@jnl{ZAp}}                 % Zeitschrift fuer Astrophysik
\def\nat{\aaref@jnl{Nature}}              % Nature
\def\aplett{\aaref@jnl{Astrophys.~Lett.}} % Astrophysics Letters
\def\apspr{\aaref@jnl{Astrophys.~Space~Phys.~Res.}} % Astrophysics Space Physics Research
\def\physrep{\aaref@jnl{Phys.~Rep.}}      % Physics Reports
\def\physscr{\aaref@jnl{Phys.~Scr}}       % Physica Scripta
\def\commat{\aaref@jnl{Comm.~Math.~Phys.}}              % Communications in Mathematical Physics
\def\science{\aaref@jnl{Science}}               % Science
\def\cqg{\aaref@jnl{Classical Quant.~Grav.}}            % Classical and Quantum Gravity
\def\jpcs{\aaref@jnl{JPCS}}                                     % Journal of Physics Conference Series
\def\ijmpd{\aaref@jnl{Int.~J.~Mod.~Phys.~D}}                    % International Journal of Modern Physics D
\def\grg{\aaref@jnl{Gen.~Relat.~Gravit.}}               % General Relativity and Gravitation
\def\rpp{\aaref@jnl{Rep.~Prog.~Phys.}}          % Reports on Progress in Physics
\def\npa{\aaref@jnl{Nucl.~Phys.~A}}        % Nuclear Physics A
\def\lrr{\aaref@jnl{Living Rev.~Rel.}}                   % Living reviews in relativity
\def\jcap{\aaref@jnl{J.~Cosmology Astropart.~Phys.}}    % Journal of cosmology and astroparticle physics
\def\rmp{\aaref@jnl{Rev.~Mod.~Phys.}}   %Reviews of modern physics
\def\epjc{\aaref@jnl{Eur.~Phys.~J.~C}}

\allowdisplaybreaks[1]
\begin{document}

\color{black}       %% For one column

\title{Cosmological constraints in symmetric teleparallel gravity with bulk viscosity}

% \author{Raja Solanki\orcidlink{0000-0001-8849-7688}}
% \email{rajasolanki8268@gmail.com}
% \affiliation{Department of Mathematics, Birla Institute of Technology and
% Science-Pilani,\\ Hyderabad Campus, Hyderabad-500078, India.}
\author{Dheeraj Singh Rana\orcidlink{0000-0002-4401-8814}}
\email{drjrana2@gmail.com}
\affiliation{Department of Mathematics, Birla Institute of Technology and
Science-Pilani,\\ Hyderabad Campus, Hyderabad-500078, India.}
\author{P.K. Sahoo\orcidlink{0000-0003-2130-8832}}
\email{pksahoo@hyderabad.bits-pilani.ac.in}
\affiliation{Department of Mathematics, Birla Institute of Technology and
Science-Pilani,\\ Hyderabad Campus, Hyderabad-500078, India.}

%%%%%%%%%%%%%%%%%%%%%%%%%%%%%%%%%%%%  DATE  %%%%%%%%%%%%%%%%%%%%%%%%%%%%%%%%%%%%

\date{\today}
\begin{abstract}

In this study, we explore the accelerated expansion of the universe within the framework of modified $f(Q)$ gravity. The investigation focus on the role of bulk viscosity in understanding the universe's accelerated expansion. Specifically, a bulk viscous matter-dominated cosmological model is considered, with the bulk viscosity coefficient expressed as $\zeta= \zeta_0 \rho H^{-1} + \zeta_1 H $. We consider the power law $f(Q)$ function $f(Q)=\alpha Q^n $, where $\alpha$ and $n$ are arbitrary constants and derive the analytical solutions for the field equations corresponding to a flat FLRW metric. Subsequently, we used the combined Cosmic Chronometers (CC)+Pantheon+SH0ES sample to estimate the free parameters of the obtained analytic solution. We conduct Bayesian statistical analysis to estimate the posterior probability by employing the likelihood function and the MCMC random sampling technique, along with the AIC and BIC statistical assessment criteria. In addition, we explore the evolutionary behavior of significant cosmological parameters. The effective equation of state (EOS) parameter predicts the accelerating behavior of the cosmic expansion phase. Further, by the statefinder and  $Om(z)$ diagnostic test, we found that our viscous model favors quintessence-type behavior and can successfully describe the late-time scenario.\\

\end{abstract}

\maketitle

\textbf{Keywords:} $f(Q)$ gravity, statefinder parameter, Equation of state parameter, and Bulk Viscosity.

\section{Introduction}\label{sec1}
\justify
General Relativity (GR) is the most successful theory to explain gravitational interactions. It has successfully completed a number of observational tests, including the solar system test, and it has predicted the existence of black holes and gravitational waves, which have been confirmed through recent observations. The standard $ \Lambda $CDM model (based on GR) is the best model to describe observed cosmological phenomena. But the model has been having trouble explaining the value of the cosmological constant $\Lambda$, the coincidence problem, and the Hubble tension. Due to these drawbacks of $\Lambda $CDM model, several modified theories of gravity and several GR-based models besides $\Lambda $CDM model have been introduced to literature to address some or all of the aforementioned problems.
Typically, gravitational modifications are constructed by introducing additional terms into the Einstein-Hilbert Lagrangian. This leads to various formulations such as $f(R)$ gravity, Gauss-Bonnet gravity, $f(G)$ gravity, $f(P)$ gravity, and Horndeski/Galileon scalar-tensor theories, among others. 
Alternatively, one can obtain torsion-based gravities such as $f(T)$, $f(T,T_G)$ and $f(T,B)$ by introducing new terms to the equivalent formulation of gravity.
However, there is a third approach to developing a new type of modified gravity. This method begins with "symmetric teleparallel gravity", which relies on the non-metricity scalar $Q$, and extends it by introducing a function $f(Q)$ in the Lagrangian.
The modified theory of f(Q) gravity yields intriguing cosmological phenomena at the background level. 
Furthermore, it has effectively analyzed against a range of observational data with regards to both background and perturbation aspects, such as the Cosmic Microwave Background (CMB) \cite{R.R.,Z.Y.}, Supernovae type Ia (SNIa) \cite{Riess,Perlmutter}, Baryonic Acoustic Oscillations (BAO) \cite{D.J.,W.J.}, growth data, etc. This analysis suggests that $f(Q)$ gravity could potentially pose a challenge to the $\Lambda$CDM scenario.\\
Initially, viscosity was introduced in cosmic fluid to study the early inflation era of cosmic evolution without assuming any kind of dark energy component. A non-causal theory of viscosity is introduced by Eckart by considering first-order deviations from equilibrium. Later, the causal theory of viscosity presented by Israel and Stewart, based on the assumption of second-order deviations from equilibrium \cite{C.E.,W.I.,W.I.-2,W.I.-3}. To study the late-time cosmic acceleration, the causal theory of viscosity has been taken into consideration. the analysis of deviations of the second order gives two distinct coefficients of viscosity, namely shear and bulk viscosity. In a homogeneous universe, the velocity gradient related to shear viscosity diminishes in the cosmic fluid. Thus, within an isotropic, homogeneous FLRW background, only the bulk viscosity coefficient holds significance. The bulk viscosity can be described as the pressure needed to restore thermal stability when the cosmic fluid expands alongside the universe's expansion. Essentially, modification of GR accounts for the cosmic expansion, while the viscosity coefficients contribute to the pressure term, propelling cosmic acceleration. In recent times, there has been considerable research into the role of bulk viscosity in cosmic evolution, as indicated by a variety of references \cite{IB-1,IB-2,IB-3,IB-4,IB-5,JM,AVS,MAT,ADD7,ADD8}.

In this study, we will present our analysis and findings within the context of the newly introduced $f(Q)$ gravity framework \cite{J.B.}. The standard GR describes spacetime with a non-zero curvature tensor while non-metricity and torsion both disappears. In contrast, the teleparallel equivalent of GR (TEGR) describes spacetime with a non-zero torsion tensor while non-metricity and curvature both disappears. Additionally, the symmetric teleparallel equivalent of GR (STEGR) describes spacetime with a non-zero non-metricity tensor while torsion and curvature both disappears. The simple extensions of GR, TEGR, and STEGR are represented by $f(R)$, $f(T)$, and $f(Q)$ gravity respectively \cite{GTG}.
An important benefit of $f(Q)$ gravity lies in its field equations, which remain second order with respect to the scale factor.  This characteristic allows for the inclusion of higher-order Lagrangians without encountering complications. An additional benefit of $f(Q)$ gravity is its automatic fulfillment of Bianchi's identity. In contrast, in $f(T)$ gravity, the existence of the anti-symmetric component poses challenges. Moreover, various extensions of $f(Q)$ theory have been introduced in the literature, including the commonly referenced $f(Q,T)$ theory \cite{m1} and the Weyl-type $f(Q,T)$ theory \cite{m2}. For more on symmetric teleparallel gravity and its extension one can follow references \cite{ADD1,ADD2,ADD3,ADD4,ADD5,ADD6}.
In this work, we investigate the significance of a well motivated viscosity parametrization under the coincident $f(Q)$ gravity formalism in order to probe the late time accelerating behavior of the expansion phase of the universe. We calculate the exact solution of the assumed model and then confront it from the cosmological observations. Further, we predict the outcomes of the investigation with the observed phenomenon and then provide a thorough comparision with the existing literature.
The structure of this manuscript is as follows. In Sec \ref{sec2}, we present the Geometry with non-metricity. In Sec \ref{sec3}, Flat FLRW Universe in Symmetric Teleparallel Cosmology with Bulk Viscous Matter is presented. In Sec \ref{sec4a}, we estimate the median value of model parameters and Bulk viscous parameters, utilizing the Cosmic Chronometer + Pantheon+ SH0ES data samples. Moreover to assess the robustness of the MCMC process, we employ AIC and BIC statistical assessment. Further in Sec \ref{sec5},  we investigate the Evolutionary Behavior of Cosmological Parameters. Lastly, in Sec \ref{sec6}, we conclude our findings.

 \section{Geometry with non-metricity}\label{sec2}
 \justify
In the context of the $f(Q)$ gravity theory, we examine a teleparallel torsion-free geometric background, which is attained by nullifying both the Riemann tensor and the torsion tensor, i.e. $R^\alpha_{\: \beta\mu\nu} = 0$ and $T^\alpha_{\ \mu\nu} = 0$. Under the restrictions of symmetric teleparallelism, a complete inertial-affine connection is provided. In any chosen gauge, the general affine connection denoted as $X^\alpha_{\ \mu\nu}$, defining tensors in parallel transport and covariant derivative, can be expressed using the following parameterization \cite{jimenez1},
\begin{equation}\label{2a}
X^\alpha \,_{\mu \beta} = \frac{\partial x^\alpha}{\partial \xi^\rho} \partial_ \mu \partial_\beta \xi^\rho
\end{equation}
Additionally, for certain gauge selections, this affine connection $X^\alpha \,_{\mu \beta}$ becomes zero. This particular gauge is commonly referred as a coincident gauge. The non-metricity tensor, which arises due to the incompatibility of the connection within the theory, is the main focus of the $f(Q)$ class of theories. The non-metricity tensor for the affine connection \eqref{2a} is defined as follows:
\begin{equation}\label{2b}
Q_{\alpha\mu\nu}\equiv\nabla_\alpha g_{\mu\nu}
\end{equation} 
Furthermore, the disformation tensor $L^\alpha_{\ \mu\nu}$ is described as the difference of generic affine connection $X^\alpha_{\ \mu\nu}$ and the Levi-Civita connection $\Gamma^\alpha_{\ \mu\nu}$, i.e., $L^\alpha_{\ \mu\nu}=X^\alpha_{\ \mu\nu}-\Gamma^\alpha_{\ \mu\nu}$.
Thus, it can be concluded that
\begin{equation}\label{2c}
L^\alpha_{\ \mu\nu}\equiv\frac{1}{2}(Q^{\alpha}_{\ \mu\nu}-Q_{\mu \ \nu}^{\ \alpha}-Q_{\nu \ \mu}^{\ \alpha})
\end{equation}
The superpotential tensor $P^\lambda\:_{\mu\nu}$ can be defined as
\begin{equation}\label{2e}
4P^\lambda\:_{\mu\nu} = -Q^\lambda\:_{\mu\nu} + 2Q_{(\mu}\:^\lambda\:_{\nu)} + (Q^\lambda - \tilde{Q}^\lambda) g_{\mu\nu} - \delta^\lambda_{(\mu}Q_{\nu)}
\end{equation}
Where $Q_\alpha = Q_\alpha\:^\mu\:_\mu \: and  \tilde{Q}_\alpha = Q^\mu\:_{\alpha\mu}$ are two traces of non-metricity tensor.
In $f(Q)$ theory, the fundamental element non-metricity scalar $Q$  explains the gravitation and non-metricity scalar $Q$  can be calculated by utilizing The superpotential tensor as 
\begin{equation}\label{2d}
Q = -Q_{\lambda\mu\nu}P^{\lambda\mu\nu} 
\end{equation}
In a symmetric teleparallelism background, i.e., a flat and symmetric connection, the action for $f(Q)$ gravity is given by \cite{jimenez2}:
\begin{equation}\label{2f}
S= \int{\frac{1}{2}f(Q)\sqrt{-g}d^4x} + \int{L_m\sqrt{-g}d^4x}
\end{equation}
where $f(Q)$ is the arbitrary function of the scalar term $Q$, $L_m$ is the Lagrangian density, and $g=det( g_{\mu\nu} )$.
The metric field equation can be obtained by 
Varying the action term \eqref{2f} with respect to the metric. The metric field equation is given as follows:
\begin{widetext}
\begin{equation}\label{2g}
\frac{2}{\sqrt{-g}}\nabla_\lambda (\sqrt{-g}f_Q P^\lambda\:_{\mu\nu}) + \frac{1}{2}g_{\mu\nu}f+f_Q(P_{\mu\lambda\beta}Q_\nu\:^{\lambda\beta} - 2Q_{\lambda\beta\mu}P^{\lambda\beta}\:_\nu) = -\mathcal{T}_{\mu\nu}
\end{equation}
\end{widetext}
Here, $\mathcal{T}_{\mu\nu}$ represents the stress-energy tensor defined as,
\begin{equation}\label{2h}
\mathcal{T}_{\mu\nu} = \frac{-2}{\sqrt{-g}} \frac{\delta(\sqrt{-g}L_m)}{\delta g^{\mu\nu}}
\end{equation}
Moreover, when hypermomentum is absent, the equation \eqref{2f} varies with respect to the connection, resulting in the following connection field equation:
\begin{equation}\label{2i}
\nabla_\mu \nabla_\nu (\sqrt{-g}f_Q P^{\mu\nu}\:_\lambda) =  0 
\end{equation}

 \section{THE f(Q) COSMOLOGICAL MODEL IN FLRW BACKGROUND WITH VISCOUS MATTER}\label{sec3}
 \justify
We assume spatial flatness, homogeneity, and isotropy in the universe and consequently use the FLRW metric to define the line element,
\begin{equation}\label{3a}
ds^2= -dt^2 + a^2(t)[dx^2+dy^2+dz^2]
\end{equation}
Here, $a(t)$ is the scale factor of the universe. Now Coincident gauge coordinates are utilized in the gauge taken into account in the line element \eqref{3a}, suggesting that the metric is the only fundamental variable.
But using a different gauge creates a generic connection, which leads to a non-trivial contribution to the field equations, via the non-metricity scalar in particular \cite{Hohmann2,fQfT1}.  As we have fixed the coincident gauge so the connection is trivial, and hence the LHS of the equation \eqref{2i} vanishes and the metric is only the fundamental variable. Since we have used diffeomorphisms to fix the coincident gauge, one could think that we are not allowed to select any particular lapse function. However, the special case of $f(Q)$ theories does allow so because $Q$ retains
a residual time-reparameterisation invariance, as already explained in \cite{J.B.} so we will use this symmetry to set lapse function $N=1$. The line element \eqref{3a} has the non-metricity scalar $Q$ as follows
\begin{equation}\label{3b}
 Q= 6H^2  
\end{equation}

In literature, the coefficient of bulk viscosity $\zeta$ is frequently presumed to depend on factors such as the expansion rate, its time derivative, and the energy density. However, generally $\zeta$ can be function of all these variables, i.e., $\zeta = \zeta (H,\dot{H},\rho)$. a new form of bulk viscosity coefficient $\zeta \sim \zeta_i H^{1-2i} \rho^i$ \cite{LGG} has been introduced in the literature recently. This particular value $i=0$ corresponds to a bulk viscosity coefficient of the form $\zeta \propto H$ that is commonly attributed to describe unified viscous models along with dissipative dark matter, whereas the case $i=1$ corresponds to a bulk viscosity coefficient of the form $\zeta \propto H^{-1}\rho$ could drive accelerated expansion \cite{LGG}. In this article, we consider following form of the bulk viscosity coefficient for our analysis \cite{ADD10},
\begin{equation}
 \zeta= \zeta_1 H + \zeta_0 \rho H^{-1}    
\end{equation}
Note that $\zeta_0$ is the dimensionless quantity whereas $\zeta_1$ has the unit $ML^{-1}$.\\
The corresponding energy momentum tensor is
\begin{equation}\label{3d}
\mathcal{T}_{\mu\nu}=(\rho+ p_v)u_\mu u_\nu + p_v g_{\mu\nu}  
\end{equation}
Here, $\rho$ stands for the matter-energy density, with $p_v = p - 3\zeta H$ as pressure due to viscosity and $u^\mu = (1,0,0,0)$ representing the four-velocity vector. In the case of bulk viscosity, the field equations are expressed as \cite{LZ}:
\begin{equation}\label{3z}
3H^{2}= \rho_{eff} = \frac{1}{2f_{Q}}\left(-\rho +\frac{f}{2}\right)  
\end{equation}
\begin{equation}
\dot{H}+3H^{2}+\frac{\dot{f_{Q}}}{f_{Q}}H=\frac{1}{2f_{Q}}\left({p_v}+
\frac{f}{2}\right) \label{3e}
\end{equation}
Furthermore, the matter conservation equation is,
\begin{equation}\label{3f}
\dot{\rho} + 3H\left(\rho+{p_v}\right)=0
\end{equation}

Since we are only focusing on the dust scenario for further analysis, the effective pressure will become ${p_v}=-3\zeta H$.
\justify We assume the following power law $f(Q)$ model \cite{RSU},  
\begin{equation}\label{3g}
f(Q)=\alpha Q^n,\ \ \ \alpha \neq 0  
\end{equation}
In particular, for $\alpha=-1$ and $n=1$, the standard Friedmann equations for GR can be obtained.
the Field equations \eqref{3z} and \eqref{3e} for the considered $f(Q)$ model \eqref{3g} becomes,
\begin{equation}\label{3i}
2\rho=\left(1-2n \right) \alpha 6^n H^{2n}  
\end{equation}
\begin{equation}\label{3j}
\dot{H} + \frac{3}{4n}{(2-3 \zeta_0)} H^2 = \frac{3 \zeta_1 H^{4-2n}}{2n(2n-1)\alpha 6^{n-1}H^{2n-2}}     
\end{equation}
The solution for this differential equation is,

\begin{multline}\label{3o}
H(z)= \Biggl\{H^{2n-2}_0(1+z)^{\frac{3 (n-1) (2-3 \zeta_0)}{2n}} \\ +\frac{2\zeta_1} {\alpha (2n-1) 6^{n-1}(2-3 \zeta_0)} \left[(1+z)^{\frac{3 (n-1) (2-3 \zeta_0)}{2n}}-1\right]\Biggl\}^{\frac{1}{2n-2}}
\end{multline}

\section{ Estimation of Parameter Using Observational Data}\label{sec4a}
\justify

In this segment, we focus on utilizing cosmic chronometer and Pantheon Supernovae datasets to estimate the values for parameters appears in the expression of $H(z)$ for our assumed model. We utilize the Markov Chain Monte Carlo (MCMC) random sampling technique alongside with Bayesian analysis and the Scipy optimization method in the Python package emcee \cite{Mackey/2013} to estimate the median values of model parameters and viscosity parameters.

\subsubsection{Cosmic Chronometer datasets}
\justify
In this article, we have used cosmic chronometer (CC) datasets which includes 31 Hubble points, obtained by using the method of differential age (DA) in the range of redshift given as $ 0.07 \leq z \leq 2.41 $. The complete list of 31 data points of the H(z) measurements utilizing DA method are listed in \cite{RS}. The $\chi^2$ function for the considered H(z) data set is given as follows:
\begin{equation}\label{4a}
\chi_{CC}^{2}=\sum\limits_{k=1}^{31}
\frac{[H_{th}(z_{k},\theta)-H_{obs}(z_{k})]^{2}}{
\sigma _{H(z_{k})}^{2}}.  
\end{equation}
In this context, $H_{obs}$ refers to the Hubble parameter value obtained from observational data. On the other hand, $H_{th}$ stands for its theoretical value calculated at $z_{k}$ within the parameter space $\theta$, and $\sigma_{H(z_{k})}$ denotes the associated error.
\subsubsection{Pantheon+SH0ES datasets}
\justify

In this current study, we analyze a recently published  revised Pantheon+SH0ES  dataset of Pantheon supernovae. The dataset comprises 1701 supernova samples, each associated with its observed distance modulus $\mu^{obs}$ in the redshift range $z \in [0.001, 2.3]$.
The Pantheon+SH0ES datasets surpass previous compilations of Type Ia supernovae (SNIa), integrating the latest observations available. In recent times, several collections of Type Ia supernova data have surfaced, such as Union \cite{R30}, Union2 \cite{R31}, Union2.1 \cite{R32}, JLA \cite{R33}, Pantheon \cite{R34}, and the most recent one, Pantheon+SH0ES \cite{R35}.

The $\chi^2$ function associated with the Pantheon+SH0ES data points is as follows:
\begin{equation}\label{4c}
\chi^2_{SN}=  D^T C^{-1}_{SN} D,
\end{equation}
Here, $C_{SN}$ denoting the covariance matrix \cite{R35} and the vector $D$ is described as $D=m_{Bi}-M-\mu^{th}(z_i)$, where $m_{Bi}$ and $M$ are apparent magnitude and the absolute magnitude respectively.
Furthermore, The theoretical value of distance
modulus is given by
\begin{equation}\label{4d}
\mu(z)= 5log_{10} \left[ \frac{D_{L}(z)}{1 Mpc}  \right]+25, 
\end{equation}
where 
\begin{equation}\label{4e}
D_{L}(z)= c(1+z) \int_{0}^{z} \frac{ dx}{H(x,\theta)}
\end{equation}
Here $\theta$ stands for parameter space.
Unlike the Pantheon dataset, the Pantheon+SH0ES dataset effectively addresses the degeneracy between the parameters $H_0$ and $M$ by redefining the vector $D$ as follows:
\begin{equation}\label{4f}
\bar{D} = \begin{cases}
     m_{Bi}-M-\mu_i^{Ceph} & i \in \text{Cepheid hosts} \\
     m_{Bi}-M-\mu^{th}(z_i) & \text{otherwise}
    \end{cases}   
\end{equation}
Here $\mu_i^{Ceph}$ independently estimated using Cepheid calibrators. Hence, the relation \eqref{4c} becomes $\chi^2_{SN}=  \bar{D}^T C^{-1}_{SN} \bar{D} $.\\

\subsubsection{CC + Pantheon+SH0ES dataset}
\justify
We have used the combined cosmic chronometer + Pantheon+SH0ES dataset to get the estimated values of the model and bulk viscous parameters. The chi-squared function for the Hubble + Pantheon+SH0ES dataset is $\chi_{total}$=$\chi_{CC}^{2}+\chi_{SN}^{2}$.

\begin{widetext}
\begin{center}
\begin{figure}[H]
\includegraphics[scale=0.6]{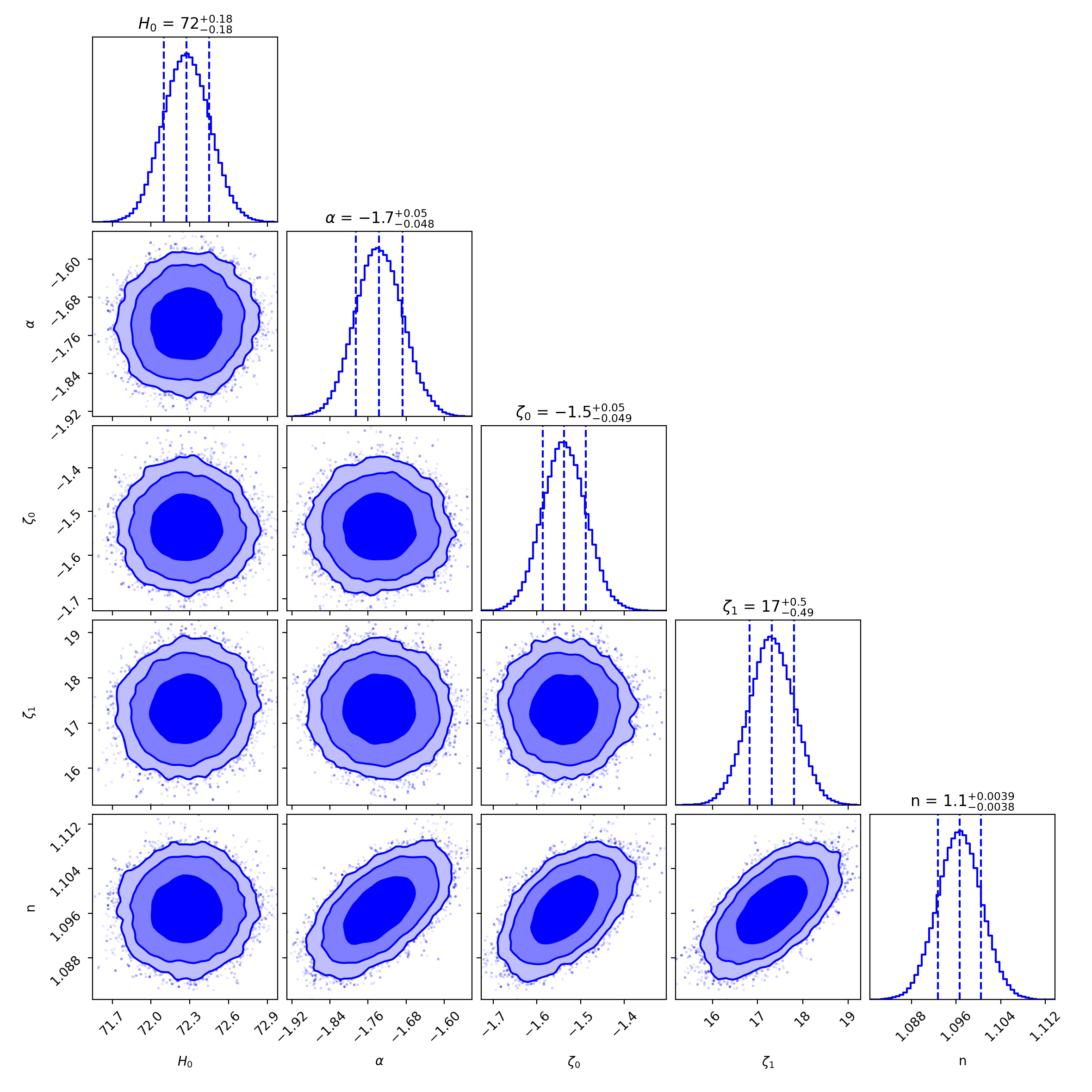}
\caption{Constraints on parameters at $1-\sigma$, $2-\sigma$ and $3-\sigma$ confidence interval using combined CC + pantheon+SH0ES data set.}\label{f1}
\end{figure}
\end{center}
\end{widetext}

The obtained estimated values of the model parameter and viscosity parameters by using Hubble + Pantheon+SH0ES data set are $H_0=72^{+0.18} _{-0.18}$, $\alpha=-1.7^{+0.05} _{-0.048}$, $\zeta_0=-1.5^{+0.05} _{-0.049}$, $\zeta_1=17^{+0.05} _{-0.049}$ ,$n=1.1^{+0.0039} _{-0.0038}$ and we got the minimum value of the ${\chi^2}_{total} $ as ${\chi^2}_{min} = 1633.683$.
\subsubsection{Model Comparison}
\justify

It is essential to conduct a statistical analysis utilizing the Akaike Information Criterion (AIC) and Bayesian Information Criterion (BIC) in order to assess the robustness of our MCMC analysis. The expression for AIC is as follows:

\begin{equation}
 AIC={\chi^2}_{min}+2d 
\end{equation}
Where $d$ indicates the number of parameters in the specified model. To compare our model with standard $\Lambda CDM$ model, we define $\Delta AIC = AIC_{Model} - AIC_{\Lambda CDM}$. value of $\Delta AIC $ less than 2 indicates substantial support for the proposed theoretical model, while Range $4 < \Delta AIC < 7 $ indicates moderate support. Furthermore, if the $ \Delta AIC $ value is greater than 10, it indicates a lack of support for the proposed model. The second criterion, BIC, can be described as follows:
\begin{equation}
 BIC={\chi^2}_{min}+dln(N) 
\end{equation}
In this context, N denotes the number of data samples utilized in the MCMC analysis. Similarly, if $\Delta BIC$ is less than 2, it indicates high support for the proposed theoretical model, and for $2 < \Delta BIC < 6 $, it indicates moderate support. The obtained values of $AIC$ and $BIC$ for the theoretical model are $AIC_{Model}=1643.683$ and $BIC_{Model}=1670.968$. Also, we got $\Delta AIC=0.515 $ and $\Delta BIC=15.856 $. Here, The $\Lambda CDM$ values of $AIC$ and $BIC$ are taken as $AIC_{\Lambda}=1644.198$ and $BIC_{\Lambda}=1655.112$.
Thus, the $\Delta AIC$ value clearly indicates strong support for the proposed theoretical $f(Q)$ model. But as is often known, a high $\Delta BIC$ value can be offset by a large number of parameters.

\section{Evolutionary Behavior of Cosmological Parameters}\label{sec5}
\justifying

We analyze the evolutionary behavior of several key cosmological parameters, including the effective equation of state (EoS), statefinder parameters, and the Om diagnostic parameter at constraints values of parameters obtained from the combined CC+Pantheon+SH0ES data set.

\begin{figure}[H]
\includegraphics[scale=0.6]{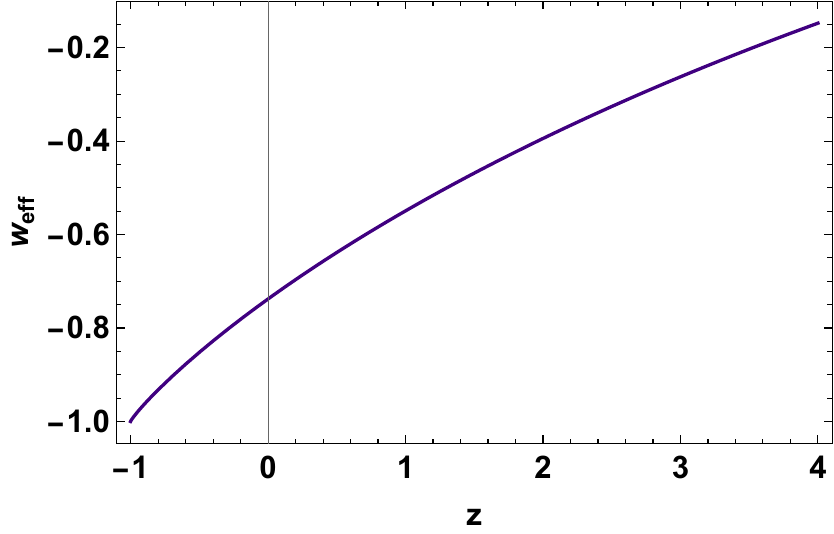}
\caption{The figure represent the behavior of the EOS parameter.}\label{f7a}
\end{figure}

The effective equation of state parameter's behavior illustrated in Figure \eqref{f7a}, indicates the accelerating expansion phase of the universe. The present value of EoS parameter for our model is obtained as $\omega_0 \approx -0.7378 $ for the combined CC + Pantheon+SH0ES sample. This obtained present value of the EoS parameter is quite consistent with the recent cosmological investigations such as Koussour et al. \cite{NN} found $\omega_0 \approx -0.756$ corresponding to power law correction in the Hubble function parametrization, whereas they found $\omega_0 \approx -0.755$ corresponding to the logarithmic correction.

The statefinder diagnostic, initially proposed by V. Sahni \cite{R37}, provides a geometric method for differentiating various dark energy models. It is based on the statefinder parameters, denoted as $r$ and $s$. These parameters are  constructed solely from the scale factor and its time derivatives. The statefinder parameters defined as follows,
 \begin{equation}\label{5a}
  r =\frac{\dddot{a}}{aH^3} \:\: \text{and} \:\: s=\frac{(r-1)}{3(q-\frac{1}{2})} 
\end{equation}

 The value $(r < 1, s > 0)$ represent the quintessence dark energy, whereas the region $(r > 1, s < 0)$ represent the phantom scenario and the value $(r = 1, s = 0)$ represent the standard $\Lambda$CDM model.

 \begin{figure}[H]
 \includegraphics[scale=0.45]{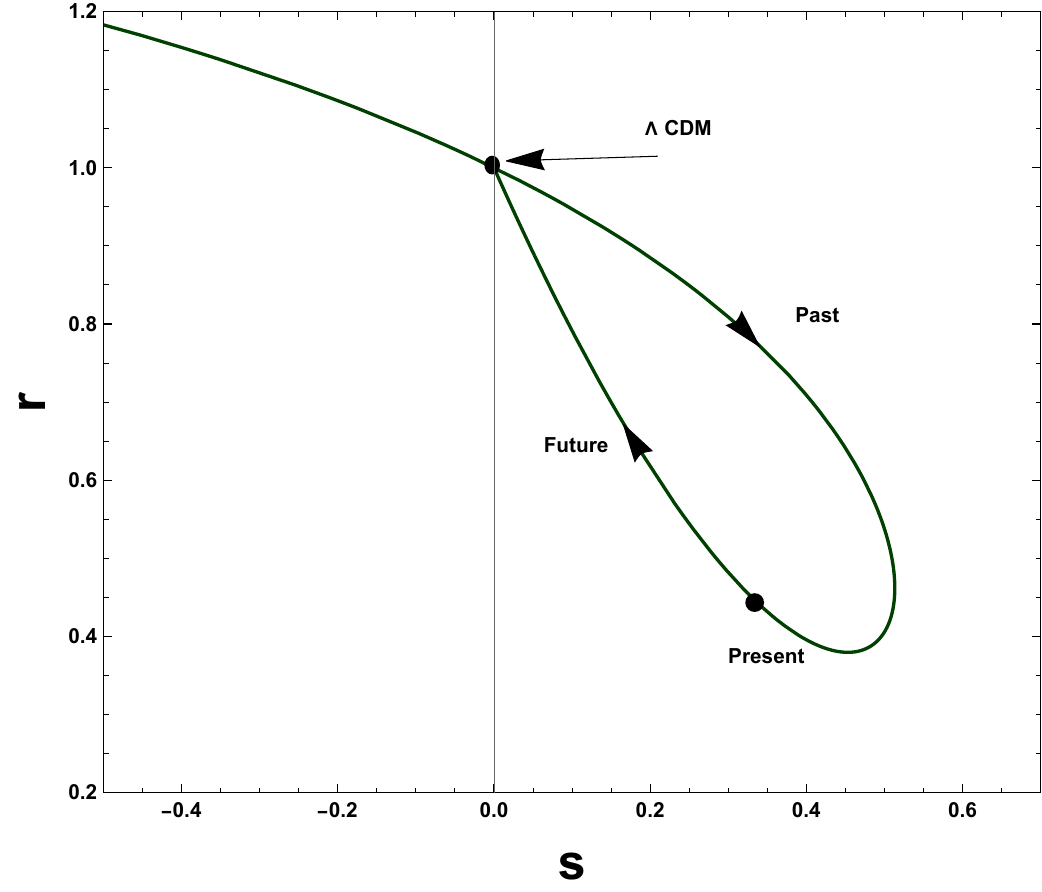}
 \caption{The figure shows the behavior of the statefinder parameters in the $r-s$ plane.}\label{f8a}
 \end{figure}

The behavior of statefinder parameters in the $r-s$ plane is presented in Fig \eqref{f8a}. The present time values of the statefinder parameters for our model is $(r_0,s_0) = (0.44,0.33)$, corresponding to the combined CC+Pantheon+SH0ES sample. Hence, the present time values of statefinder parameters of the considered viscous model favors quintessence-type behavior.

The Om diagnostic serves as a simple testing method that depends only on the cosmic scale factor's first-order derivative. For a spatially flat universe, its expression is as follows \cite{R38}:
\begin{equation}\label{5b}
 Om(z)= \frac{\big(\frac{H(z)}{H_0}\big)^2-1}{(1+z)^3-1}
 \end{equation}

The slope of the $Om(z)$ curve can tell us about the behavior of the assumed viscous model. If slope of the $Om(z)$ curve is descending throughout the domain, it indicates quintessence-like behavior of the model. while the ascending slope of the $Om(z)$ curve indicates the phantom behavior of the model. On the other hand, a constant $Om(z)$ represents the $\Lambda$CDM model.
 
\begin{figure}[H]
 \includegraphics[scale=0.48]{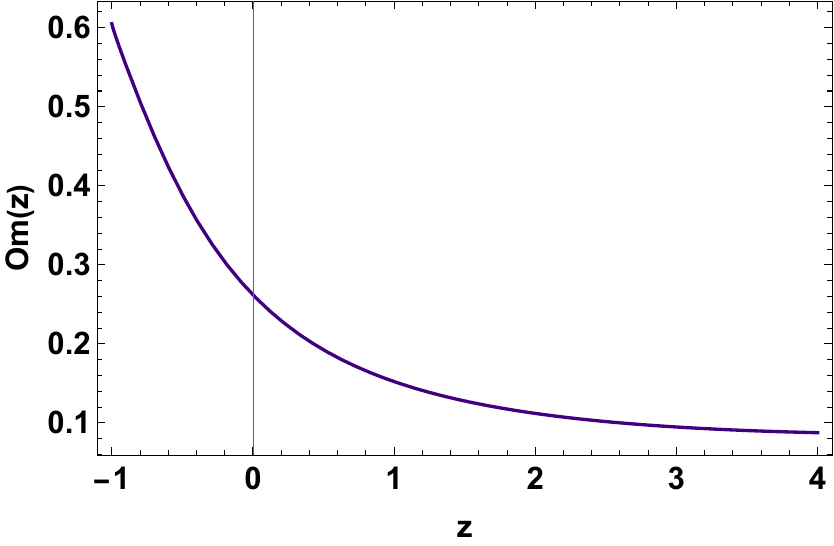}
 \caption{The figure displays the behavior of the Om diagnostic parameter}\label{f9a}
 \end{figure}

The behavior of the $Om(z)$ parameter for the assumed viscus $f(Q)$ model is represented in Fig. \eqref{f9a}. The slope of the $Om(z)$ curve is descending throughout the domain for the estimated values of parameters. Therefore, on the basis of Om diagnostic test, we can say that our viscous fluid cosmological model shows quintessence-like behavior.

\section{Conclusion} \label{sec6}
\justifying

 The study of $f(Q)$ gravity has gained substantial attention from cosmologists, encompassing a range of topics, including wormholes, black holes, late-time acceleration, inflation, etc. Meanwhile, cosmological models involving viscous fluids have gained significant interest, particularly for their description of the universe's early stages and offering insights into late-time expansion. Considering hydrodynamics, incorporating the influence of viscosity in the cosmic fluid is a reasonable assumption, given that a perfect fluid is ultimately an idealization. In this study, we explored the significance of viscosity coefficients in explaining the observed cosmic acceleration in the $f(Q)$ gravity background.

We begin with the power law $f(Q)$ model $f(Q)=\alpha Q^n $, where $\alpha$ and $n$ are arbitrary constants, along with the fluid part incorporating the coefficient of bulk viscosity $\zeta= \zeta_0 \rho H^{-1} + \zeta_1 H $. The analytical solutions of the corresponding field equations for a flat FLRW environment are presented in equation \eqref{3o}. The free parameters of the obtained solutions have been constrained using the CC+Pantheon+SH0ES sample.
We performed the Bayesian statistical analysis to estimate the posterior probability utilizing the likelihood function and the MCMC random sampling technique. We have obtained estimated values of the model parameter and viscous parameters by using CC+Pantheon+ sample. In addition, the contour plots for the free parameters $H_0, \alpha, \zeta_0$, $\zeta_1$ and $n$ within the $1\sigma-3\sigma$ confidence interval are presented in Figs \eqref{f1}. Moreover, we performed a statistical analysis utilizing the Akaike Information Criterion (AIC) and Bayesian Information Criterion (BIC) in order to assess the robustness of our MCMC analysis.  We obtained $\Delta AIC=0.515 $ and $\Delta BIC=15.856 $.
Thus, the $\Delta AIC$ value clearly indicates strong support for the proposed theoretical $f(Q)$ model, whereas a high $\Delta BIC$ value can be offset by a large number of parameters due to the presence of a factor d in multiplication of $ln(N)$. 
We also examined the evolutionary behavior of some prominent cosmological parameters. The effective equation of state parameter predicts the accelerating behavior of the expansion phase of the universe (see Fig \eqref{f7a}. The value of the EoS parameter at the present redshift $(z=0)$ is $\omega_0 \approx -0.7378 $ corresponding to the combined CC+Pantheon+SH0ES sample.
 
Further, Fig \eqref{f8a} illustrate the behavior of the assumed viscous $f(Q)$ model in the $r-s$ plane. The corresponding present time value of the statefinder parameters for our model is $(r_0,s_0) = (0.44, 0.33)$ corresponding to the model parameter constraints obtained by the combined CC+Pantheon+SH0ES sample. Hence, the statefinder parameters  favor quintessence-type behavior. Lastly, in Fig \eqref{f9a}, we illustrated the behavior of the $Om(z)$ curve, which represents a consistent negative slope across the entire domain for our assumed model. Thus, it can be inferred that our assumed viscous fluid $f(Q)$ models embodies quintessence-like behavior and can successfully describe the late-time scenario. In this article, we updated the constraints on bulk viscous and model parameters utilizing Pantheon+SH0ES samples along with CC sample, whereas the existing literature \cite{ADD3} considered only Hubble and BAO samples but not the supernovae samples, whereas the references \cite{ADD6} utilized old Pantheon samples. Moreover, the $H_0$ constraint corresponding to the Hubble, Pantheon, and the joint samples obtained in the references \cite{ADD5,ADD6} are almost identical (nearly $H_0 \approx 67 \: km/s/Mpc$) that is not consistent, since the discrepancy in the $H_0$ value for Hubble and Pantheon sample is well known in the literature known as the $H_0$ tension. In the present manuscript, we obtained $H_0 \approx 72 \: km/s/Mpc$ for the joint analysis that is in less tension. For the future work, one can thoroughly investigate the $\sigma_8$ and $S8$ tension. \\

\textbf{Data availability:} This article does not include any new data.

\section*{Acknowledgments} 
 DSR acknowledges UGC, New Delhi, India for providing Senior Research Fellowship with (NTA-UGC-Ref.No.: 211610106591). PKS  acknowledges the Science and Engineering Research Board, Department of Science and Technology, Government of India for financial support to carry out Research Project No.: CRG/2022/001847. We are very much grateful to the honorable referees and to the editor for the
illuminating suggestions that have significantly improved our work in terms of research quality, and presentation.

%%%%%%%%%%%%%%%%%%%%%%%%%%%%%%%%%%%%%%%%%%%%%%%%%%%%%%%%%%%%%%%%%%%%%%%%%%%%%%%%%
%%
%%
 

\begin{thebibliography}{90}
   
\bibitem{R.R.} R.R. Caldwell, M. Doran, \textit{Phys. Rev. D} \textbf{69}, 103517 (2004).

\bibitem{Z.Y.} Z.Y. Huang et al., \textit{JCAP} \textbf{0605}, 013 (2006).

\bibitem{Riess} A.G. Riess et al., \textit{Astron. J.} \textbf{116}, 1009 (1998). 

\bibitem{Perlmutter} S. Perlmutter et al., \textit{Astrophys. J.} \textbf{517}, 565 (1999).

\bibitem{W.J.} W.J. Percival at el., \textit{Mon. Not. R. Astron. Soc.} \textbf{401}, 2148 (2010).

\bibitem{D.J.} D.J. Eisenstein et al., \textit{Astrophys. J.} \textbf{633}, 560 (2005).

\bibitem{C.E.} C. Eckart, \textit{Phys. Rev.} \textbf{58}, 919(1940).

 \bibitem{W.I.} W. Israel, J. M. Stewart, \textit{Phys. Lett. B} \textbf{58},
213 (1976).

 \bibitem{W.I.-2} W. Israel, \textit{Ann. Phys.} (N.Y.) \textbf{100}, 310
(1976).

 \bibitem{W.I.-3} W. Israel, J. M. Stewart, \textit{Proc. R. Soc. Lond. B} 
 \textbf{365}, 43 (1979).

 \bibitem{IB-1} I. Brevik, \textit{Entropy} \textbf{2012(14)}, 2302-2310 (2012).

 \bibitem{IB-2} I. Brevik and O. Gron, \textit{Astrophys. Space Sci.} \textbf{347}, 399 (2013).

 \bibitem{IB-3} I. Brevik et al., \textit{Int. J. Mod. Phys. D} \textbf{26}, 1730024 (2017).

 \bibitem{IB-4} I. Brevik, A. N. Makarenko, and A. V. Timoshkin, \textit{Int. J. Geom. Methods Mod.} \textbf{16}, 1950150 (2019)

 \bibitem{IB-5} I. Brevik and B. D. Normann, \textit{Symmetry} \textbf{2020(12)}, 1085 (2020).

 \bibitem{JM} N. D. J. Mohan, A. Sasidharan, and T. K. Mathew, \textit{Eur. Phys. J. C} \textbf{77}, 849 (2017).

 \bibitem{AVS} A. V. Astashenok, S. D. Odintsov, and A. S. Tepliakov, \textit{Nucl. Phys. B} \textbf{974}, 115646 (2022).

 \bibitem{MAT} A. Sasidharan and T. K. Mathew, \textit{Eur. Phys. J. C} \textbf{75}, 348 (2015).

\bibitem{ADD7} J Satish and R Venkateswarlu, \textit{Chin. J. Phys.} \textbf{54}, 830-838 (2016).

\bibitem{ADD8} P. S. Debnath,  \textit{Int. J. Geom. Methods Mod. Phys.} \textbf{16}, 1950005 (2019).

\bibitem{J.B.} J. B. Jim\'enez et al., \textit{Phys. Rev. D} \textbf{98}, 044048 (2018).

\bibitem{GTG} J. B. Jim\'enez, L. Heisenberg, and T. Koivisto, \textit{Universe}, \textbf{5}, 173 (2019).

\bibitem{m1} Y. Xu et al., \textit{Eur. Phys. J. C} \textbf{79}, 708 (2019).

\bibitem{m2} Y. Xu et al., \textit{Eur. Phys. J. C} \textbf{80}, 449 (2020).

\bibitem{ADD1} O. Sokoliuk1 and A. Baransky, \textit{Astron. Nachr.} \textbf{343}, 220003 (2022). 

\bibitem{ADD2} A. Dixit, D. C. Maurya, and A. Pradhan, \textit{Int. J. Geom. Methods Mod. Phys.} \textbf{19}, 12 (2022).

\bibitem{ADD3} M. M. Gohain and K. Bhuyan, \textit{Phys. Dark Univ.} \textbf{43}, 101424 (2024).

\bibitem{ADD4} M. Koussour et al., \textit{Chin. J. Phys.} \textbf{90}, 97-107 (2024).

\bibitem{ADD5} M. Koussour et al., \textit{Phys. Dark Univ.} \textbf{45}, 101527 (2024).

\bibitem{ADD6} M. Koussour et al., \textit{Mod. Phys. Lett. A} \textbf{39}, 2450023 (2024).

\bibitem{jimenez1} J. B. Jim\'enez et al., \textit{Phys. Rev. D} {\bf 101}, 103507(2020).

\bibitem{jimenez2} J.B. Jim\'enez, L. Heisenberg, and T. S. Koivisto, \textit{JCAP} {\bf 1808}, 039 (2018).


\bibitem{Hohmann2} M. Hohmann,  \textit{Phys. Rev. D} \textbf{104}, 124077 (2021). 

\bibitem{fQfT1} F. D'Ambrosio, L. Heisenberg, S. Kuhn, \textit{Class. Quantum Grav.} \textbf{39}, 025013 (2022).

\bibitem{LGG} L. Gomez, G. Palma, A. Rincon, N. Cruz, E. Gonzalez, \textit {Eur. Phys. J. Plus} \textbf{138}, 738 (2023).

\bibitem{ADD10} V. A Pai and T. K. Mathew, \textit{Class. Quantum Grav.} \textbf{41}, 31 (2024).

\bibitem{LZ} R. Lazkoz et al., \textit{Phys. Rev. D} \textbf{100}, 104027 (2019).



\bibitem{RSU} R. Solanki et al., \textit{Universe} \textbf{9(1)}, 12 (2023).

\bibitem{Mackey/2013} D. F. Mackey et al., \textit{Publ. Astron. Soc. Pac.} \textbf{125}, 306(2013). 

 \bibitem{RS} R. Solanki et al., {\it Phys. Dark Univ.} {\bf 32}, 100820 (2021).






\bibitem{R30} M. Kowalski et al., \textit{Astrophys. J.} \textbf{686}, 749-778 (2008).

\bibitem{R31} R. Amanullahet al., \textit{Astrophys. J.} \textbf{716}, 712-738 (2010).

\bibitem{R32} N. Suzuki et al., \textit{Astrophys. J.} \textbf{746}, 85, (2012).

\bibitem{R33} M. Betoule et al., \textit{Astron. Astrophys.} \textbf{568}, A22 (2014).
 
\bibitem{R34} D. M. Scolnic et al., \textit{Astrophys. J.} \textbf{859}, 101 (2018).

\bibitem{R35} D. M. Scolnic et al., \textit{Astrophys. J.} \textbf{938}, 113 (2022). 

\bibitem{NN} M. Koussour et al., \textit{Eur. Phys. J. Plus} \textbf{139}, 179 (2024).


 % \bibitem{Tom} T. Ortin, \textit{Gravity and Strings}, Cambridge Monographs on Mathematical Physics (Cambridge University Press (2015). 

 % \bibitem{CMS} C. Misner et al., \textit{Gravitation}, Princeton University Press (2017). 

 % \bibitem{JLL} J.B. Jim\'enez, L. Heisenberg, and T. S. Koivisto, \textit{JCAP} {\bf 1808}, 039 (2018).

 

 % \bibitem{LAR} L. Heisenberg, M. Hohmann, and S. Kuhn, \textit{arXiv}, arXiv:2212.14324 (2022).

 % \bibitem{Ryden} B. Ryden, \textit{Introduction to Cosmology} (Addison Wesley, SanFrancisco, United States of America, 2003).

% \bibitem{FDA} F. D\'Ambrosio, L. Heisenberg, and S. Kuhn, \textit{Class. Quantum Grav.} \textbf{39}, 025013 (2021).



% \bibitem{ZHH} Z. Hassan, S. Mandal, and P. K. Sahoo, \textit{Forts. Phys.} \textbf{69}, 2100023 (2021). 

% \bibitem{DZ1} A. Sasidharan and T. K. Mathew, \textit{J. High Energy Phys.} \textbf{06}, 138 (2016).

% \bibitem{DZ2} J. Ren and X. H. Meng, \textit{Phys. Lett. B} \textbf{633}, 1 (2006).

\bibitem{R37} V. Sahni et al., {\it JETP Lett.} {\bf 77}, 201 (2003).

\bibitem{R38} V. Sahni, A. Shafieloo, and A.A. Starobinsky, \textit{Phys. Rev. D} \textbf{78}, 103502 (2008).



% \bibitem{Planck} Planck Collaboration, \textit{Astron. Astrophys.} \textbf{641}, A6 (2020).  

 % \bibitem{Scolnic/2018} D.M. Scolnic et al., \textit{Astrophys. J.} \textbf{859}, 101(2018).

 % \bibitem{BMS} R. Kessler and D. Scolnic, \textit{Astrophys. J.} \textbf{836},56 (2017).

% \bibitem{BAO1} C. Blake et al., \textit{ Mon. Not. Roy. Astron. Soc.} \textbf{418}, 1707 (2011).

% \bibitem{BAO2} W. J. Percival et al., \textit {Mon. Not. Roy. Astron. Soc.} \textbf{401}, 2148 (2010).
 
% \bibitem{BAO6} R. Giostri et al.,\textit{ J. Cosm. Astropart. Phys.} \textbf{1203}, 027 (2012).

% \bibitem{V.S.} V. Sahni et al., {\it JETP Lett.} {\bf 77}, 201 (2003).


\end{thebibliography}
\end{document}